# On the local integrability of almost-product structures defined by space-time metrics

D. H. Delphenich
Spring Valley, OH, USA

**Abstract:** The splitting of the tangent bundle of space-time into temporal and spatial sub-bundles defines an almost-product structure. In particular, any space-time metric can be locally expressed in "time-orthogonal" form, in such a way that whether or not that almost-product structure is locally generated by a coordinate chart is a matter of the integrability of the Pfaff equation that the temporal 1-form of that expression for the metric defines. When one applies that analysis to the known exact solutions to the Einstein field equations, one finds that many of the common ones are completely-integrable, although some of the physically-interesting ones are not.

**1. Introduction.** – One of the unavoidable aspects of the extension from three dimensions to four that the shift from non-relativistic physics to relativistic physics entails is that one must still have some way of getting back to three dimensions, since that is essentially where one makes all of one's observations and measurements. As Max Born once pointed out, any measurement is made in the rest space of the measuring device, so when one ultimately wishes to verify a theoretical prediction in a laboratory or observatory, one must first account for the mathematical form that the measurements or observations will take in order to evaluate them numerically.

The most common way of getting back from four dimensions to three in the theory of relativity is to split the tangent spaces $T_x M$ to the space-time manifold $M$ into direct sums $[\mathbf{t}]_x \oplus \Sigma_x$ of time-like lines $[\mathbf{t}]_x$ and space-like hyperplanes $\Sigma_x$. Such a splitting of $T(M)$, which is also called a *Whitney sum* decomposition, is called an *almost-product* structure on $M$, since if $M$ were indeed a product manifold of the form $\mathbb{R} \times \Sigma$ then the splitting of the tangent bundle would follow naturally from the product structure by differentiation; i.e., $[\mathbf{t}] = T(\mathbb{R})$, $\Sigma(M) = T(\Sigma)$. Hence, in order to go from an almost-product structure to a product structure, one must integrate, which implies that one must address the degrees of integrability of the differential systems that are defined by the sub-bundles $[\mathbf{t}]$ and $\Sigma(M)$. The former differential system is a line field, and therefore completely-integrable, but the latter has codimension-one, so its integrability is more involved. The author has examined this issue is several other papers [**1**], in which one will also find a more complete set of references to the work of others.

As explained in Møller [**2**], any space-time metric $g$ can be locally expressed in "time-orthogonal" form, namely, $g = (\theta^0)^2 - \gamma_s$, where $\theta^0$ is a time-like 1-form whose annihilating hyperspaces define the spatial hyperplanes $\Sigma_x$ in each tangent space, and $\gamma_s$ is a (Riemannian) metric on those spatial hyperplanes; the temporal line $[\mathbf{t}]_x$ is then generated by the vector that is metric dual to the 1-form $\theta^0$. The process of getting from $g_{\mu\nu} \, dx^\mu \, dx^\nu$ to the latter form is essentially a process of "completing the square." Since the dual spatial hyperplanes in the cotangent spaces $\Sigma_x^*$ can be locally spanned by the holonomic coframe $dx^i$, $i = 1, 2, 3$, the only issue in regard to the integrability of the splitting is the integrability of the exterior differential system $\theta^0 = 0$, which defines what was always called a "Pfaff equation." The theory of the degree of integrability of that equation exists in a useful form by now [1], and can be applied to the special case in question. More specifically, one can examine how it works for some of the many exact solutions of the Einstein field equations for gravitation.

The basic structure of this article is therefore entirely straightforward: In the second section, we define the concept of an almost-product structure and the ways that one can define such a thing in practice. In particular, we address $1 + (n - 1)$ splittings of tangent bundles, which are the ones that are of immediate interest. In section **3**, we discuss the basic theory of the integrability of the Pfaff equation. In section **4**, we introduce a few elementary terms from hydrodynamics in order to illustrate the basic constructions in the theory of the Pfaff equation. In section **5**, we discuss the way that one can locally complete the square for any Lorentzian metric $g$ on space-time in order to put it into "time-orthogonal" form and then address the degree of integrability of the Pfaff equation that the temporal 1-form $\theta^0$ defines. In section **6**, we apply that basic analysis to numerous common exact solutions of the Einstein equations. In the final section **7**, we recapitulate the goal and results of our analysis.

## 2. Almost-product structures.

**2. Almost-product structures.** – When an $n$-dimensional differentiable manifold $M$ takes the form of a product manifold $M_1 \times M_2$, there will be a corresponding direct-sum decomposition [2] $T(M_1) \oplus T(M_2)$ of the tangent bundle $T(M)$ that is defined by all sums of tangent vectors $\mathbf{v}_1 \in T_x(M_1)$ with tangent vectors in $\mathbf{v}_2 \in T_y(M_2)$ to produce tangent vectors $\mathbf{v}_1 + \mathbf{v}_2 \in T_{(x, y)}(M_1 \times M_2)$. Hence, every point $(x, y) \in M_1 \times M_2$ has a tangent space $T_{(x, y)}(M_1 \times M_2)$ that splits into a direct sum $T_{(x, y)}(M_1) \oplus T_{(x, y)}(M_2)$ of subspaces that are isomorphic to $T_{(x)}(M_1)$ and $T_{(y)}(M_2)$, respectively. In particular, they have complementary dimensions, so $n = n_1 + n_2$.

*a. Definition and basic properties.* – If one has only a Whitney sum decomposition of a tangent bundle $T(M) = \Sigma_1(M) \oplus \Sigma_2(M)$ then it will not necessarily follow that there will be a corresponding product structure to the manifold $M$. One then says that such a decomposition defines an *almost-product structure*. However, one can say that the sub-bundles $\Sigma_1(M)$ and $\Sigma_2(M)$ do define transverse differential systems on $M$; i.e., at every point of $M$ there are linear subspaces in the tangent space that are complementary to each other and intersect at the origin.

Hence, a first step towards testing the limits to which $M$ admits a true product structure would be to examine the degree of integrability of each of the differential systems. At best, one might hope for transverse foliations of $M$ by integral submanifolds of $\Sigma_1(M)$ and $\Sigma_2(M)$; i.e., every pair of intersecting integral submanifolds would intersect transversely (viz., their tangent spaces would be transverse).

---

[1]   The author has produced a survey of applications of the theory of the Pfaff problem to physics in the form of [**3**], which also includes references to more established works on the Pfaff problem and its applications.

[2]   Recall that when one has a direct-sum decomposition of a vector space $V = V_1 \oplus V_2$, one can express any vector $\mathbf{v} \in V$ uniquely as a sum $\mathbf{v}_1 + \mathbf{v}_2$ with $\mathbf{v}_1 \in V_1$ and $\mathbf{v}_2 \in V_2$.

Naturally, a Whitney sum decomposition $T(M) = \Sigma_1(M) \oplus \Sigma_2(M)$ will also imply a corresponding Whitney sum decomposition $T^*(M) = \Sigma_1^*(M) \oplus \Sigma_2^*(M)$. The latter decomposition amounts to the normal-space decomposition $N_1(M) \oplus N_2(M)$ that is dual to $\Sigma_1(M) \oplus \Sigma_2(M)$, except that $N_1(M) = \Sigma_2^*(M)$ and $N_2(M) = \Sigma_1^*(M)$.

Whenever one has an almost-product structure $T(M) = \Sigma_1(M) \oplus \Sigma_2(M)$ on a manifold $M$, one can define a corresponding decomposition of all tensor products of the bundles $T(M)$ and $T^*(M)$ by using the distributivity of the tensor product. For instance, when one looks at only tensor powers of $T^*(M)$, such as one considers for space-time metric tensors and exterior differential forms, one will get successive decompositions ([1]):

$$\Sigma_1^* \oplus \Sigma_2^*, \qquad (\Sigma_1^* \otimes \Sigma_1^*) \oplus (\Sigma_1^* \otimes \Sigma_2^*) \oplus (\Sigma_2^* \otimes \Sigma_1^*) \oplus (\Sigma_2^* \otimes \Sigma_2^*), \qquad \ldots$$

In particular, one sees that the total covariant tensor algebra ([2]):

$$T^{0,*}(M) = \mathbb{R} \oplus T^* \oplus (T^* \otimes T^*) \oplus (T^* \otimes T^* \otimes T^*) \oplus \ldots$$

includes contributions from the total covariant tensor algebra:

$$T^{0,*}(\Sigma_1) = \mathbb{R} \oplus \Sigma_1^* \oplus (\Sigma_1^* \otimes \Sigma_1^*) \oplus (\Sigma_1^* \otimes \Sigma_1^* \otimes \Sigma_1^*) \oplus \ldots$$

and the total covariant tensor algebra:

$$T^{0,*}(\Sigma_2) = \mathbb{R} \oplus \Sigma_2^* \oplus (\Sigma_2^* \otimes \Sigma_2^*) \oplus (\Sigma_2^* \otimes \Sigma_2^* \otimes \Sigma_2^*) \oplus \ldots,$$

as well as tensor products of a mixed nature, such as $\Sigma_1^* \otimes \Sigma_2^*$.

When one projects to completely-symmetric tensors, one will get:

$$S^*(M) = \mathbb{R} \oplus [\Sigma_1^* \oplus \Sigma_2^*] \oplus [(\Sigma_1^* \odot \Sigma_1^*) \oplus (\Sigma_1^* \odot \Sigma_2^*) \oplus (\Sigma_2^* \odot \Sigma_2^*)] \oplus \ldots,$$

which includes contributions from the completely-symmetric covariant tensor algebras $S^*(\Sigma_1)$ and $S^*(\Sigma_2)$, along with mixed symmetric tensor products. When one projects to completely-antisymmetric tensors, one will get:

$$\Lambda^*(\Sigma_1) = \mathbb{R} \oplus [\Sigma_1^* \oplus \Sigma_2^*] \oplus [(\Sigma_1^* \wedge \Sigma_1^*) \oplus (\Sigma_1^* \wedge \Sigma_2^*) \oplus (\Sigma_2^* \wedge \Sigma_2^*)] \oplus \ldots,$$

which includes contributions from the completely-antisymmetric covariant tensor algebras $\Lambda^*(\Sigma_1)$ and $\Lambda^*(\Sigma_2)$, along with mixed exterior products.

---

([1])  When there is no question of what base manifold we are dealing with, we shall abbreviate the notation for the bundle accordingly.

([2])  A good reference for multilinear algebra, as it relates to the present discussion, is Greub [**4**].

For example, if $g$ is a metric tensor field on $M$ (of any signature type) and one has a splitting of $T(M)$ into $\Sigma_1 \oplus \Sigma_2$ then one can decompose $g$ into:

$$g = g_{11} + g_{12} + g_{22},$$

in which the summands are defined by starting with decompositions $\mathbf{v} = \mathbf{v}_1 + \mathbf{v}_2$ and $\mathbf{w} = \mathbf{w}_1 + \mathbf{w}_2$, such that:

$$g_{11}(\mathbf{v}, \mathbf{w}) = g(\mathbf{v}_1, \mathbf{w}_1),$$
$$g_{12}(\mathbf{v}, \mathbf{w}) = g(\mathbf{v}_1, \mathbf{w}_2) + g(\mathbf{v}_2, \mathbf{w}_1),$$
$$g_{22}(\mathbf{v}, \mathbf{w}) = g(\mathbf{v}_2, \mathbf{w}_2).$$

Similarly, if $F$ is a 2-form then there will be a corresponding decomposition:

$$F = F_{11} + F_{12} + F_{22},$$

with:

$$F_{11}(\mathbf{v}, \mathbf{w}) = F(\mathbf{v}_1, \mathbf{w}_1),$$
$$F_{12}(\mathbf{v}, \mathbf{w}) = F(\mathbf{v}_1, \mathbf{w}_2) + F(\mathbf{v}_2, \mathbf{w}_1),$$
$$F_{22}(\mathbf{v}, \mathbf{w}) = F(\mathbf{v}_2, \mathbf{w}_2).$$

*b. Defining almost-product structures.* – There are two ways of defining any $k$-dimensional linear subspace $A$ of an $n$-dimensional vector space $V$ ([1]):

1. One can span it directly by way of a basis of vectors $\{\mathbf{e}_i, i = 1, \ldots, k\}$.

2. One can annihilate it by way of $n - k$ covectors $\{\theta^a, a = 1, \ldots, n - k\}$.

The corresponding constructions for the tangent bundle $T(M)$ and cotangent bundle $T^*(M)$ of a manifold $M$ are more topologically sensitive, since one not only has to contend with the possible non-existence of globally-non-zero vector fields and globally-non-zero covector fields, but the non-existence of $k$ globally-non-zero vector fields that will span $k$-dimensional subspaces of the tangent spaces at each point of $M$. Basically, one is up against the "degree of parallelizability" of $M$. Since we shall be more oriented towards the existing constructions of general relativity theory, which are generally local constructions that start with coordinate systems, we shall not go further in the topological direction, and will take advantage of the fact that $T(M)$ and $T^*(M)$ will be trivial bundles over some neighborhood of each point of $M$, and in particular, over every coordinate neighborhood.

When $T(M)$ and $T^*(M)$ are trivial, one can always span them with an *n-frame field*, i.e., $n$ linearly-independent vector fields $\{\mathbf{e}_i, i = 1, \ldots, n\}$, and an *n-coframe field*, i.e., $n$ linearly-independent covector fields $\{\theta^i, i = 1, \ldots, n\}$, resp. If $T = \Sigma_1 \oplus \Sigma_2$ (so one also has $T^* = \Sigma_1^* \oplus \Sigma_2^*$ ) then the frame field $\{\mathbf{e}_i, i = 1, \ldots, n\}$ will be called *adapted* to the splitting when $\{\mathbf{e}_i, i = 1, \ldots, k\}$ spans $\Sigma_1$, and $\{\mathbf{e}_a, a = k + 1, \ldots, n\}$ spans $\Sigma_2$; dually, $\{\theta^i, i = 1, \ldots, k\}$ spans $\Sigma_1^*$ and $\{\theta^a, a = k + 1, \ldots, n\}$ spans $\Sigma_2^*$. One can also say that the coframe field is adapted to the splitting when $\{\theta^a, a = k + 1, \ldots, n\}$ annihilates $\Sigma_1$ and $\{\theta^i, i = 1, \ldots, k\}$ annihilates $\Sigma_2$.

---

([1])   In the time of Julius Plücker, these two ways of describing a subspace corresponded to regarding it as a "locus" and an "envelope," respectively. In projective geometry, that took the form of generating the (projective) subspace as the join of $k$ points (in general position) or the meet of $n - k$ hyperplanes.

We shall find that it is often convenient to sometimes span $\Sigma_1$ with $\{\mathbf{e}_i, i = 1, \ldots, k\}$ and annihilate $\Sigma_2$ with $\{\theta^i, i = 1, \ldots, k\}$. In particular, that is the case for a $1 + (n - 1)$ splitting, for which the one-dimensional tangent space of $\Sigma_1$ can be spanned by a single non-zero vector field $\mathbf{e}_0$ and the codimension-one vector space $\Sigma_2$ can be annihilated by a single non-zero covector field $\theta^0$. The transversality condition will then take the form of saying that one must have $\theta^0(\mathbf{e}_0) \neq 0$, since the vanishing of that expression would imply that $\mathbf{e}_0$ would be contained in the hyperplane that is annihilated by $\theta^0$.

When a metric $g$ has been defined on $T(M)$, one can substitute orthogonality for transversality. Hence, it will suffice to define one of the two bundles $\Sigma_1$ and $\Sigma_2$ so that the other one can be defined to be its orthogonal complement. Once can also speak of restricting the spanning $k$-frame $\{\mathbf{e}_i, i = 1, \ldots, k\}$ and the annihilating $k$-coframe $\{\theta^i, i = 1, \ldots, k\}$ to orthonormal frames and coframes, respectively.

When one is given an almost-product structure $T = \Sigma_1 \oplus \Sigma_2$, one can also decompose the differential operators $d$ and $d_\wedge$ [1] into components that relate to differentiation with respect to the two sub-bundles:

$$d = dx^i \otimes \partial_i + dx^a \otimes \partial_a, \tag{2.1}$$

$$d_\wedge = dx^i \wedge \partial_i + dx^a \wedge \partial_a. \tag{2.2}$$

For instance, the differentials of functions $f$ and vector fields $\mathbf{v} = v^\mu \partial_\mu$ take the forms:

$$df = \partial_i f \, dx^i + \partial_a f \, dx^a$$

$$d\mathbf{v} = \partial_i v^\mu \, dx^i \otimes \partial_\mu + \partial_a v^\mu \, dx^a \otimes \partial_\mu,$$

and if $\alpha = \alpha_\mu \, dx^\mu$ is a 1-form then:

$$d_\wedge\alpha = \tfrac{1}{2}(\partial_i \alpha_\mu - \partial_\mu \alpha_i) \, dx^i \wedge dx^\mu + \tfrac{1}{2}(\partial_a \alpha_\mu - \partial_\mu \alpha_a) \, dx^a \wedge dx^\mu.$$

Note that since the index $\mu$ goes through the entire range from 1 to $n$, the expressions for $d\mathbf{v}$ and $d_\wedge\alpha$ can be decomposed further into:

$$d\mathbf{v} = \partial_i v^j \, dx^i \otimes \partial_j + \partial_i v^a \, dx^i \otimes \partial_a + \partial_a v^i \, dx^a \otimes \partial_i + \partial_a v^b \, dx^a \otimes \partial_b$$

$$d_\wedge\alpha = \tfrac{1}{2}(\partial_i \alpha_j - \partial_j \alpha_i) \, dx^i \wedge dx^j + (\partial_i \alpha_a - \partial_a \alpha_i) \, dx^i \wedge dx^a + \tfrac{1}{2}(\partial_a \alpha_b - \partial_b \alpha_a) \, dx^a \wedge dx^b.$$

---

[1]   $d_\wedge$ is our notation for the exterior derivative operator, so as not to confuse it with the $d$ that represents the differential of a differentiable map.

Hence, one cannot generally decompose $d$ and $d_\wedge$ into two pieces $d_1$ and $d_2$ ($d_{1\wedge}$ and $d_{2\wedge}$, resp.) that behave like the differential operators on a product manifold, but one must include components that represent the mixing of field components from one bundle with coordinates from the other. For $d\mathbf{v}$, one might denote them by $d_{12}$ and $d_{21}$, while the corresponding mixed operator for $d_\wedge \alpha$ might be called $d_{12\wedge}$. Of course, the number of mixed differential operators will depend upon the rank of the tensor field in question.

*c.* 1 + (*n*–1) *splittings.* – The case in which the direct-sum decomposition of each tangent space $[\mathbf{t}]_x \oplus \Sigma_x$ consists of a line $[\mathbf{t}]_x$ and a transverse hyperplane $\Sigma_x$ is particularly tractable. That is because the integrability of the line field $[\mathbf{t}]$ into a congruence of curves is guaranteed, while the integrability of the hyperplane field $\Sigma$ comes down to the theory of the Pfaff equation when one represents the hyperplanes as annihilating hyperplanes for a projective family $[\theta^0]$ of 1-forms.

A 1 + (*n*–1) decomposition $[\mathbf{t}] \oplus \Sigma$ can then be characterized by the pair $\{\mathbf{t}, \theta^0\}$, together with the transversality condition that $\theta^0(\mathbf{t}) \neq 0$. When $T(M)$ also has a metric $g$ that is defined on it, one can also strengthen transversality to orthogonality and make the 1-form $\theta^0$ the metric-dual to the vector field $\mathbf{t}$:

$$\theta^0 = i_{\mathbf{t}}\, g \qquad [\text{i.e., } \theta^0(\mathbf{v}) = g(\mathbf{t}, \mathbf{v}) \text{ for all } \mathbf{v}]. \tag{2.3}$$

In that case, the transversality condition will take the form:

$$\theta^0(\mathbf{t}) = \|\,\mathbf{t}\,\|^2. \tag{2.4}$$

In the case of the Lorentzian structure on space-time, one usually imposes the further requirement that the vector field $\mathbf{t}$ must be time-like, so $\|\,\mathbf{t}\,\|^2 > 1$. Indeed, $\mathbf{t}$ will usually take the form of a time-like four-velocity vector field for the motion of some object that gets referred to as an "observer." One might imagine an inertial measurements unit for an aircraft or ship as a specific example of an observer, since it contains an orthogonal triad of linear accelerometers for the measurement of linear accelerations and another orthogonal triad of rate gyros for the measurement of angular velocities. We shall refer to $\theta^0$ as a *temporal 1-form* and its annihilating hyperplane as a *spatial hyperplane.*

The forms that the decomposition of the operators $d$ and $d_\wedge$ take in the particular case of a 1 + (*n*–1) splitting are:

$$d = c\, dt \otimes \partial_t + dx^i \otimes \partial_i\,, \qquad d_\wedge = c\, dt \wedge \partial_t + dx^i \wedge \partial_i\,, \tag{2.5}$$

so for a vector field $\mathbf{v} = v^\mu\, \partial_\mu = (1/c)\, v^0\, \partial_t + v_i\, dx^i$ and a 1-form $\alpha = \alpha_\mu\, dx^\mu = c\, \alpha_0\, dt + \alpha_i\, dx^i$ one will have:

$$d\mathbf{v} = c^2\, \partial_t\, v^0\, dt \otimes \partial_t + c\, \partial_t\, v^i\, dt \otimes \partial_i + c\, \partial_i\, v^0\, dx^i \otimes \partial_t + \partial_i\, v^j\, dx^i \otimes \partial_j\,, \tag{2.6}$$

$$d_\wedge \alpha = -\, c\, \partial_i\, \alpha_0\, dt \wedge dx^i + \tfrac{1}{2}(\partial_i\, \alpha_j - \partial_j\, \alpha_i)\, dx^i \wedge dx^j\,, \tag{2.7}$$

respectively.

We shall reserve the notations $d_s$ and $d_{s^\wedge}$ for the operators that behave as if one were dealing with a true product manifold; i.e., if:

$$t = t^{i_1 \cdots i_r}_{j_1 \cdots j_s} \, dx^{j_1} \otimes \cdots \otimes dx^{j_s} \otimes \partial_{i_1} \otimes \cdots \otimes \partial_{i_r} \qquad (2.8)$$

is a tensor, and:

$$\alpha = \tfrac{1}{k!} \alpha_{i_1 \cdots i_k} \, dx^{i_1} \wedge \cdots \wedge dx^{i_k} \qquad (2.9)$$

is a $k$-form then:

$$d_s t = \partial_k t^{i_1 \cdots i_r}_{j_1 \cdots j_s} \, dx^k \otimes dx^{j_1} \otimes \cdots \otimes dx^{j_s} \otimes \partial_{i_1} \otimes \cdots \otimes \partial_{i_r} \qquad (2.10)$$

and

$$d_{s^\wedge} \alpha = \tfrac{1}{k+1} \partial_{[j} \alpha_{i_1 \cdots i_k]} \, dx^j \wedge dx^{i_1} \wedge \cdots \wedge dx^{i_k}, \qquad (2.11)$$

in which the [...] in the indices refers to complete antisymmetrization.

**3. Integrability of differential systems.** – We shall now discuss the main mathematical technique that will be applied in what follows.

*a. Basic definitions.* – A *differential system* of dimension $k$ (codimension $n - k$) on an $n$-dimensional differentiable manifold $M$ is a smooth vector sub-bundle $\mathfrak{D}^k(M)$ of $T(M)$ of rank $k$ everywhere. That is, it is a smooth association of a $k$-dimensional vector subspace $\mathfrak{D}_x(M)$ of $T_x(M)$ with every point $x \in M$. (For some authors, that arrangement is called a *distribution*.)

A differential system $\mathfrak{D}^k(M)$ is said to have *degree of differentiability* $m$ ($\leq k$) if there is a foliation of $M$ by $m$-dimensional leaves such that their tangent spaces are subspaces of the subspaces of $\mathfrak{D}^k(M)$ at each point. Hence, $M$ will be partitioned into a disjoint sum of $m$-dimensional submanifolds, and every point of $M$ has a chart around it such that the coordinates of a point on each leaf of the foliation that contains the point will take the form $(x^1, \ldots, x^m, x_0^{m+1}, \ldots, x_0^n)$, where $x_0^{m+1}, \ldots, x_0^n$ are constants. $\mathfrak{D}^k(M)$ is *completely integrable* iff its degree of differentiability is $k$.

*Frobenius's theorem* says that a differential system $\mathfrak{D}^k(M)$ is completely integrable iff it is *involutive:* That is, if we let $\mathfrak{X}(M)$ represent the (infinite-dimensional) vector subspace of the vector space $\mathfrak{X}(\mathfrak{D})$ of all vector fields on $M$ that is defined by vector fields that take their values in the subspaces of $\mathfrak{D}^k(M)$ then the Lie algebra that is generated by the vector fields in $\mathfrak{X}(\mathfrak{D})$ must be $\mathfrak{X}(\mathfrak{D})$ itself; i.e., the Lie bracket of any two vector fields in $\mathfrak{X}(\mathfrak{D})$ must be a vector field in $\mathfrak{X}(\mathfrak{D})$.

A useful corollary to Frobenius's theorem is that any one-dimensional differential system must be completely integrable. Now, a one-dimensional differential system is a global line field on $M$, so its integral submanifolds will be integral curves that have those lines for their tangents. (A one-dimensional foliation is usually referred to as a *congruence of curves* in relativity.) One should be careful to distinguish between the complete integrability of one-dimensional differential systems and the existence of local flows for non-zero vector fields, since every non-zero vector field generates a line field. The difference is that a flow imposes a common parameterization for all of the integral curves, while a one-dimensional foliation does not specify the parameterizations of the integral curves, only their existence.

Since the congruence of integral curves that is associated with the line field [**t**] does not come with a canonical curve parameter, in general, the way that one must choose one is to choose a vector field **t** that will serve as an exemplar for the line field [**t**] and integrate the autonomous system of ordinary differential equations:

$$\frac{dx}{d\tau} = \mathbf{t}\,(x(\tau)) \tag{3.1}$$

that **t** defines.

When one chooses a local coordinate chart $(U, x^i)$ and its natural frame field, one can express **t** in the form $t^i \partial_i$, and the latter system will take the form:

$$\frac{dx^i}{d\tau} = t^i(x^j(\tau))\,. \tag{3.2}$$

With a different parameterization $s = s\,(\tau)$ for the curves, so one must have $ds / d\tau \neq 0$, and the latter system will take the form:

$$\frac{dx^i}{ds} = \frac{d\tau}{ds}t^i(x^j(s))\,, \tag{3.3}$$

which means that the vector field **t** can be replaced with the vector field $(d\tau/ds)\,\mathbf{t}$, which will still generate the same line field [**t**], since the scalar factor is non-zero.

*b. Exterior differential systems.* – The most common way of defining differential systems is by way of *exterior differential systems.*   That is, the *k*-dimensional tangent subspaces of $\mathfrak{D}^k(M)$ are annihilated by $(n - k)$ 1-forms $\{\theta^a, a = 1, \ldots, n - k\}$, which one expresses as the system of equations:

$$\theta^a = 0 \qquad (a = 1, \ldots, n - k). \tag{3.4}$$

More precisely, this is a system of linear algebraic equations at every point $x \in M$ whose solution will be the linear subspace $\mathfrak{D}_x(M)$ at that point.

However, one can turn that system of equations into a system of (generally partial) differential equations when one looks at what it means for a submanifold $\sigma : S \to M$, where $S$ is *m*-dimensional, to be an integral submanifold.  The condition that $\sigma$ should be integral is that:

$$\sigma^* \theta^a = 0 \qquad (a = 1, \ldots, n - k), \tag{3.5}$$

in which $\sigma^*$ refers to the pull-back map $\sigma^* : \Lambda^*(M) \to \Lambda^*(S)$ that is associated with $\sigma$; hence, in this case, if $p \in S$, and $\mathbf{v} \in T_pS$ then:

$$(\sigma^* \theta^a)_p(\mathbf{v}) = \theta^a_{\sigma(p)}(d\sigma\big|_p(\mathbf{v})). \qquad (3.6)$$

In order to turn this into differential equations, one considers a local coordinate chart $(U, x^i)$ about $x \in M$ and another one $(V, u^a)$ about $p \in S$. The function $\sigma$ will then become a system of $n$ equations in $m$ independent variables:

$$x^i = \sigma^i(u^a). \qquad (3.7)$$

If the 1-form $\theta^a$ is expressed in terms of the natural coframe field $dx^i$ as:

$$\theta^a = \theta^a_i dx^i \qquad (3.8)$$

then its pull-back $\sigma^* \theta^a$ will take the form:

$$\sigma^* \theta^a = \theta^a_i \frac{\partial \sigma^i}{\partial u^\alpha} du^\alpha; \qquad (3.9)$$

i.e., one replaces all $dx^i$ with $(\partial \sigma^i / \partial u^\alpha) du^\alpha$. $\sigma^* \theta^a$ will vanish iff all of its components vanish, and that will yield the system of $(n-k) \times m$ first-order partial differential equations in the $n$ functions $\sigma^i(u^\alpha)$ of $m$ independent variables:

$$\theta^a_i \frac{\partial \sigma^i}{\partial u^\alpha} = 0. \qquad (3.10)$$

In the case of complete integrability, the solutions will be $n$ functions of $k$ independent variables.

*b. The Pfaff problem.* – A 1-form $\theta$ is sometimes referred to as a *Pfaffian form*, and the exterior equation $\theta = 0$ is then referred to as the *Pfaff equation*. The *Pfaff problem* is then to find all solutions to the exterior differential system that is defined by the Pfaff equation; i.e., to find its degree of integrability. What we have posed in (3.4) is typically referred to as a *Pfaff system*, and the study of the integrability of Pfaff systems is much more involved than the integrability of a single Pfaff equation ([1]). However, we shall be concerned with only a single Pfaff equation, so we shall discuss only the simpler case.

An equivalent form for Frobenius's theorem when the differential system $\mathfrak{D}^{n-1}(M)$ is the codimension-one system that is defined by the Pfaff equation for a 1-form $\theta$ is that $\mathfrak{D}^{n-1}(M)$ is completely integrable iff:

$$\theta \wedge d_\wedge \theta = 0. \qquad (3.11)$$

---

([1])   That problem goes back to the work of Élie Cartan [**5**] and Erich Kähler [**6**]; see also [**7**] for a more modern discussion.

Now, when Cartan [**8**] discussed the Pfaff problem (see also Goursat [**9**]), he showed that this condition is only one of a series of conditions that determine the degree of integrability of the Pfaff problem for θ. One first defines a series of $k$-forms in terms of θ and $d_\wedge θ$:

$$θ, \ d_\wedge θ, \ θ \wedge d_\wedge θ, \ d_\wedge θ \wedge d_\wedge θ, \ θ \wedge d_\wedge θ \wedge d_\wedge θ, \ \ldots$$

Of course, since one cannot have non-zero $k$-forms for $k > n$ ($n$ = the dimension of $M$), this sequence must truncate for any finite-dimensional manifold.

The vanishing of the first $k$-form in the sequence (viz., θ) gives the Pfaff equation in question. The vanishing of the second one $d_\wedge θ$ would say that θ is closed. Locally (or globally, if $M$ were simply-connected), that would also make it exact, so there would be a 0-form $f$ such that:

$$θ = d\phi. \tag{3.12}$$

The integral submanifolds would then be the level hypersurfaces of the function $\phi$, so the Pfaff equation would be completely-integrable.

The difference between the vanishing of $d_\wedge θ$ and the vanishing of $θ \wedge d_\wedge θ$ is that one can accomplish the latter when:

$$θ = λ \, d\phi \tag{3.13}$$

for some pair of non-zero 0-forms $λ$, $\phi$; traditionally, $λ$ was called an "integrating factor." As long as $λ$ is everywhere non-zero, one will see that the vanishing of θ is equivalent to the vanishing of $d\phi$, so the integral submanifolds will again be level hypersurfaces of $\phi$.

The next step up from having θ take the form $λ \, d\phi$ is for it to take the form:

$$θ = d\phi + λ \, d\mu \,. \tag{3.14}$$

That would make:

$$θ \wedge d_\wedge θ = (d\phi + λ \, d\mu) \wedge d λ \wedge d\mu = d\phi \wedge dλ \wedge d\mu, \tag{3.15}$$

which does not generally vanish, although:

$$d_\wedge θ \wedge d_\wedge θ = dλ \wedge d\mu \wedge dλ \wedge d\mu = 0. \tag{3.16}$$

This time, the integral submanifolds will be the intersections of the level surfaces of $\phi$ and those of $\mu$, so their dimension will have been reduced by one, and the Pfaff equation will no longer be completely integrable.

The next form that θ can take is:

$$θ = ν \, d\phi + λ \, d\mu \,. \tag{3.17}$$

In this case:

$$d_\wedge θ \wedge d_\wedge θ = (dν \wedge d\phi + dλ \wedge d\mu) \wedge (dν \wedge d\phi + dλ \wedge d\mu) = 2 \, dν \wedge d\phi \wedge dλ \wedge d\mu, \tag{3.18}$$

which does not generally vanish, but:

$$\theta \wedge d_\wedge\theta \wedge d_\wedge\theta = 2(\nu \, d\phi + \lambda \, d\mu) \wedge d\nu \wedge d\phi \wedge d\lambda \wedge d\mu = 0. \qquad (3.19)$$

The integral submanifolds in this case are still the intersections of the level hypersurfaces of $\phi$, $\mu$, since the functions $\nu$, $\lambda$ both serve as integrating factors. Hence, the degree of integrability is the same as in the previous case.

Since we shall be dealing with a four-dimensional manifold, we shall not need to go to higher dimensions. However, in general one has:

1. $\theta$ can take either of two *normal forms:*

$$\theta = \sum_{i=1}^{p} \lambda_i \, d\mu^i \quad \text{or} \qquad \theta = d\phi + \sum_{i=1}^{p} \lambda_i \, d\mu^i . \qquad (3.20)$$

2. The minimum number $2p$ ($2p + 1$, resp.) is called the *rank* of the Pfaff equation; i.e., the number of independent functions in the normal form.

3. The integral submanifolds in the first (second, resp.) case of (3.20) are of the form:

$$\mu^i = \text{const.} \qquad\qquad (\phi = \text{const.}, \, \mu^i = \text{const.}, \text{ resp.}) \qquad (3.21)$$

4. If $q$ is the number of linearly-independent differentials in the normal form for $\theta$ then the degree of integrability will be equal to $n - q$.

5. If the first vanishing $k$-form in the integrability sequence has degree $k = k_0$ then degree of integrability will be equal to $n - k_0 / 2$ when $k_0$ is even and $n - (k_0 - 1) / 2$ when $k_0$ is odd.

We summarize the results of this section in Table 1.

Table 1.  Integrability of a Pfaffian form $\theta$.

| Normal form | First non-zero $k$-form | $q$ | $k_0$ | Degree of integrability |
|---|---|---|---|---|
| $d\phi$ | $d_\wedge\theta$ | 1 | 2 | $n-1$ |
| $\mu \, d\nu$ | $\theta \wedge d_\wedge\theta$ | 1 | 3 | $n-1$ |
| $d\phi + \mu \, d\nu$ | $d_\wedge\theta \wedge d_\wedge\theta$ | 2 | 4 | $n-2$ |
| $\mu_1 \, d\nu_1 + \mu_2 \, d\nu_2$ | $\theta \wedge d_\wedge\theta \wedge d_\wedge\theta$ | 2 | 5 | $n-2$ |

**4. Some elementary fluid mechanics**. – When the vector field **t** is defined on a spatially-extended world-tube, rather than simply a world-line, a common way of specifying the vector field **t** in more detail is to treat it as the flow velocity vector field **v** of a moving mass of deformable matter ([1]).

*a. Streamlines. Path-lines. Flow tubes* [**10-14**]. – In non-relativistic continuum mechanics, the flow velocity $\mathbf{v}(t, x^i)$ of a continuum motion is defined on a region $D$ in space $\Sigma$, and can therefore the vector field **v** can vary in time, as can the region $D$ itself, such as the region that is occupied by a cloud in windy weather. For the case of fluid motion in a channel or volume with fixed walls, it is only **v** that can vary then. In the event that **v** does not vary in time, one will be dealing with *steady flow*.

In the case of steady flow, the integral curves of **v** in $D$ will not change in time, and one calls them *streamlines*. In the event that the flow is unsteady, one calls them *path-lines*. Since the spatial region $D$ is foliated by a congruence of integral curves, one can think of it as having a tubular sort of character, and one then refers to it as a *flow tube*. However, whether or not $D$ can actually be expressed as a product manifold of a cylindrical form $[a, b] \times S$ or a toral form $S^1 \times S$ for some surface $S$ is a question of integrability.

In the relativistic case, since $\mathbf{v}(t, x^i)$ is always a function of time, the distinction between streamlines and path-lines is moot, so one collectively refers to them as streamlines. Furthermore, one usually expects that the four-dimensional world-tube that is foliated by streamlines will project to spatial flow tubes that are foliated by the spatial projections of the streamlines.

*b. Vorticity.* – When one starts with a (non-relativistic) covelocity 1-form $v = v_i \, dx^i$, the exterior derivative:

$$\omega = d \wedge v = \tfrac{1}{2}(v_{i,j} - v_{j,i}) \, dx^i \wedge dx^j \tag{4.1}$$

is called the *kinematical vorticity*, or when one does not intend to distinguish it from the "dynamical vorticity," which is the exterior derivative of the momentum 1-form, simply the *vorticity*.

Of particular interest is the relativistic vorticity that one obtains from a covelocity 1-form that takes the local form:

$$u = u_0 \, dx^0 + u_i \, dx^i = \beta \, (c \, dt + v) \qquad (\beta = \left(1 - \frac{v^2}{c^2}\right)^{-1/2}, \, v_i = u_i \, / \, \beta). \tag{4.2}$$

One gets:

$$\Omega = (\partial_t u_i - c \, \partial_i \, \beta) \, dt \wedge dx^i + \tfrac{1}{2}(u_{i,j} - u_{j,i}) \, dx^i \wedge dx^j, \tag{4.3}$$

which can also be written as:

$$\Omega = d\beta \wedge (c \, dt + v) + \beta \, d \wedge v. \tag{4.4}$$

If one expands $d \wedge v$ into temporal and spatial contributions then one will get:

---

([1]) Although it is traditional to refer to the flow in hydrodynamical terms, actually, many other forms of extended matter can define a flow velocity vector field by their motion. For instance, imagine a water-balloon throwing contest, in which the water balloon will deform as it moves through the air, without being precisely a fluid.

$$d_\wedge v = dt \wedge a + \omega, \qquad (a \equiv \partial_t v), \tag{4.5}$$

which includes both the non-relativistic acceleration 1-form $a$ and the non-relativistic vorticity 2-form $\omega$ that was defined above.

The relativistic correction to the non-relativistic kinematics then comes from two places: the usual factor of $\beta$ that multiplies $d_\wedge v$ and the first term in (4.4), which is:

$$d\beta \wedge (c \, dt + v) = dt \wedge (\partial_t \beta \, v - c \, d_s \beta) + d_s \beta \wedge v.$$

Hence, the non-relativistic acceleration $a$ goes to:

$$a_r = \beta \, a + \partial_t \beta \, v - c \, d_s \beta = \partial_t u - c \, d_s \beta, \tag{4.6}$$

which agrees with the first term in (4.3), and the non-relativistic vorticity $\omega$ goes to:

$$\omega_r = \beta \, \omega + d_s \beta \wedge v = d_{s\wedge} u_s, \tag{4.7}$$

which is consistent with the second term in (4.3).

If one starts with the expression for $\beta$ in (4.2) and performs the differentiations then one will get:

$$\partial_t \beta = \frac{\beta^3}{c^2} < v, a >, \qquad d_s \beta = \frac{\beta^3}{c^2} < v, d_s v > = \frac{\beta^3}{c^2} (v^j \, \partial_i v_j) \, dx^i. \tag{4.8}$$

The first one will vanish for uniform circular motion, while the second term in the relativistic expression for $\omega$ will vanish whenever the (spatial) gradient of $\beta$ is collinear with the spatial flow velocity.

Whenever the covelocity 1-form $v$ is exact – e.g., $v = d\phi$ – one calls a choice of $\psi$ a *stream function* or *velocity potential*. Clearly, a necessary condition for that is that the vorticity must vanish; when the space (space-time, resp.) manifold is simply-connected, that condition is also sufficient. The level surfaces (hypersurfaces, resp.) of $\phi$ are then equipotentials.

Since $v(\mathbf{v}) = g(\mathbf{v}, \mathbf{v}) = v^2 \neq 0$, $\mathbf{v}$ is not contained the plane that is annihilated by $v$, and if $\mathbf{w}$ is annihilated by $v$ then $v(\mathbf{w}) = g(\mathbf{v}, \mathbf{w}) = 0$. That is, $\mathbf{v}$ is orthogonal to all vectors in the annihilating plane of $v$. When $v = \lambda \, d\phi$, one says that $\mathbf{v}$ is *surface-orthogonal*, namely it is orthogonal to the level surfaces of $\phi$. Analogous statements are true for the relativistic flow velocity $\mathbf{u}$ and covelocity $u$. Hence:

**Theorem:**

*A flow velocity vector field $\mathbf{v}$ ($\mathbf{u}$, resp.) is (hyper)surface-orthogonal iff its covelocity $v$ ($u$, resp.) defines a completely-integral exterior differential system.*

*c. Vorticity vector field. Vortex tubes.* – As long as the space manifold $\Sigma$ or space-time manifold $M$ has a volume element $V_s$ on $T(\Sigma)$ [$V$ on $T(M)$, resp.], one can associate a vector field $\boldsymbol{\omega}$ with the 2-form $\omega = d_s{}^{\wedge}v$ or $\Omega = d{}^{\wedge}u$ in two ways that depend upon the dimension of space or space-time. One then calls $\boldsymbol{\omega}$ the *vorticity vector field.*

In the non-relativistic case of three-dimensions, the spatial volume element $V_s$ on $T(\Sigma)$ is a globally non-zero 3-form:

$$V_s = dx^1 \wedge dx^2 \wedge dx^3 = \tfrac{1}{3!}\varepsilon_{i_1 i_2 i_3} dx^{i_1} \wedge dx^{i_2} \wedge dx^{i_3} , \tag{4.9}$$

which defines a sequence of linear isomorphisms from $k$-vector fields to $3-k$-forms ($k = 0, 1, 2, 3$) $\#_s : \Lambda_k \to \Lambda^{3-k}$, $\mathbf{A} \mapsto \#_s\mathbf{A}$, in which:

$$\#_s\mathbf{A} = i_{\mathbf{A}}V_s ; \tag{4.10}$$

one calls these isomorphisms the *Poincaré isomorphisms.*

For instance, a vector field $\mathbf{v} = v^i \, \partial_i$ will go to a 2-form $i_{\mathbf{v}}V_s = \tfrac{1}{2} v_{ij} dx^i \wedge dx^j$ whose components are:

$$v_{ij} = \varepsilon_{ijk} v^k. \tag{4.11}$$

More to the immediate point, the inverse transformation $\#_s^{-1}$ will take the 2-form $\omega$ to the vector field $\boldsymbol{\omega}$, whose components are ([1]):

$$\omega^i = \tfrac{1}{2} \varepsilon^{ijk} \omega_{ij} = \tfrac{1}{2} \varepsilon^{ijk} (\partial_i v_j - \partial_j v_i) = \tfrac{1}{2} (\text{curl } \mathbf{v})^i. \tag{4.12}$$

In the relativistic case, where $M$ is four-dimensional, $V$ is a globally non-zero 4-form:

$$V = dx^0 \wedge dx^1 \wedge dx^2 \wedge dx^3 = \tfrac{1}{4!}\varepsilon_{\mu_0 \cdots \mu_3} dx^{\mu_0} \wedge \cdots \wedge dx^{\mu_3} , \tag{4.13}$$

so the Poincaré isomorphisms will take the form $\# : \Lambda_k \to \Lambda^{4-k}$, $\mathbf{A} \mapsto \#\mathbf{A}$, in which:

$$\#\mathbf{A} = i_{\mathbf{A}}V. \tag{4.14}$$

In particular, a vector field $\mathbf{v} = v^\mu \, \partial_\mu$ will go to a 3-form:

$$\#\mathbf{v} = i_{\mathbf{v}}V = \tfrac{1}{3!} v_{\lambda\mu\nu} dx^\lambda \wedge dx^\mu \wedge dx^\nu, \tag{4.15}$$

whose components will be:

$$v_{\lambda\mu\nu} = \varepsilon_{\kappa\lambda\mu\nu} v^\kappa. \tag{4.16}$$

Since the inverse isomorphism $\#^{-1}$ would take a 2-form to a bivector field and a 3-form to a vector field, in order to take the relativistic vorticity 2-form $\Omega$ to a vector field, one must promote it to a 3-form somehow. The traditional way of doing that ([2]) is to use $u \wedge \Omega = u \wedge d{}^{\wedge}u$, which is also the Frobenius 3-form that relates to the complete integrability of the exterior differential system $u = 0$, as well as the "Pauli-Lubanski" spin tensor of relativistic rotational mechanics.

---

([1])   This definition of the non-relativistic vorticity vector field then differs from the more conventional ones (e.g., Lamb [**10**]) by the factor of 1/2.

([2])   Some standard references on relativistic hydrodynamics are [**12-14**].

The Poincaré-dual vector field:

$$\boldsymbol{\omega} = \#^{-1}\,(u \wedge \Omega) = \omega^{\kappa}\,\partial_{\kappa} \tag{4.17}$$

will have the components:

$$\omega^{\kappa} = \tfrac{1}{3!}\,\varepsilon^{\kappa\lambda\mu\nu}\,u_{\lambda}\,\partial_{\mu}\,u_{\nu}\,. \tag{4.18}$$

As a vector field, $\boldsymbol{\omega}$ will define an autonomous system of (three or four) ordinary differential equations by way of:

$$\frac{dx}{ds} = \boldsymbol{\omega}(x(s)) \qquad \left( \frac{dx^{A}}{ds} = \omega^{A}(x^{B}(s)), \quad A = i \text{ or } \mu \right), \tag{4.19}$$

and its integral curves will be called *vortex lines*.

One finds that in the three-dimensional case:

$$v(\boldsymbol{\omega}) = <\mathbf{v},\,\boldsymbol{\omega}> = \mathbf{V}_{s}(v \wedge \#_{s}\boldsymbol{\omega}) = \mathbf{V}_{s}(v \wedge d_{s}\wedge v), \tag{4.20}$$

in which $\mathbf{V}_{s}$ is the volume element on $T^{*}(\Sigma)$ that is dual to $V_{s}$, in the sense that $V_{s}\,(\mathbf{V}_{s}) = 1$. Hence:

**Theorem:**

*The non-relativistic flow velocity vector field* $\mathbf{v}$ *is orthogonal to the vorticity vector field* $\boldsymbol{\omega}$ *iff the flow is surface-orthogonal. That is, iff the exterior differential system $v = 0$ is completely integrable.*

In the relativistic case, one will have:

$$u\,(\boldsymbol{\omega}) = <\mathbf{u},\,\boldsymbol{\omega}> = \mathbf{V}(u \wedge \#\boldsymbol{\omega}) = \mathbf{V}(u \wedge u \wedge d\wedge u) = 0. \tag{4.21}$$

Hence:

**Theorem:**

*The relativistic flow velocity vector field* $\mathbf{u}$ *is always orthogonal to the vorticity vector field* $\boldsymbol{\omega}$ *(The exterior differential system $u = 0$ does not have to be completely integrable.)*

Just as the congruence of integral curves of the flow velocity vector field $\mathbf{v}$ (i.e., pathlines or streamlines) are thought to fill up a spatial tube that is bounded by the time evolution of the spatial extent of the mass density distribution (namely, a flow tube), similarly, the congruence of vortex lines fills up a spatial tube that one calls a *vortex tube*.

**5. Local 1 + 3 splittings defined by space-time metrics.** – Suppose $M$ is the space-time manifold, which we assume to be four-dimensional and real, and to admit a Lorentzian structure that takes the form of a symmetric, second-rank, covariant tensor field $g$ that defines a scalar product on each tangent space that makes it isometric to four-dimensional Minkowski space $\mathfrak{M}^4 = (\mathbb{R}^4, \eta)$ with the sign convention that $\eta = \text{diag}[+1, -1, -1, -1]$. In particular, for any time-like tangent vector $\mathbf{v}$, one will have that $g(\mathbf{v}, \mathbf{v}) = \|\mathbf{v}\|^2$ is positive.

A local frame field $\{\mathbf{e}_\mu, \mu = 0, 1, 2, 3\}$ is defined to be *Lorentzian* iff it is orthonormal for $g$ in the Minkowski sense:

$$g_{\mu\nu} = g(\mathbf{e}_\mu, \mathbf{e}_\nu) = \eta_{\mu\nu}. \tag{5.1}$$

Every point of $M$ admits a neighborhood on which such an orthonormal frame field exists. However, the process of orthonormalizing a linear frame to a Lorentzian frame is more involved than the Gram-Schmidt process, which could introduce division by zero for light-like vectors.

One also has that every point of $M$ has a coordinate chart $(U, x^\mu)$ on a neighborhood of that point, which then defines a natural local coframe field by way of $\{dx^\mu, \mu = 0, 1, 2, 3\}$. That will allow one to express $g$ in the general local form:

$$g = g_{\mu\nu}(x) \, dx^\mu \, dx^\nu. \tag{5.2}$$

However, it is not true that every point has a coordinate chart for which the natural coframe field is Lorentzian. That is actually a question of the integrability of $G$-structures [**15**]. (In the present case, $G = SO^+(3, 1)$, the proper orthochronous Lorentz group.) In particular, one would need the Riemannian curvature of the Levi-Civita connection that is defined by $g$ to vanish. Hence, if one deals with coordinate charts then one must usually accept that the associated natural (i.e., holonomic) coframe field will be non-Lorentzian, whereas the Lorentzian local coframe fields will be *anholonomic; i.e.,* one cannot integrate any of them into the natural coframe field for any coordinate chart.

Since most space-time metrics (i.e., solutions of the Einstein field equations) tend to be expressed in the form (5.2), we shall then use that form as the basis for a local 1+3 splitting of $T(M)$. We assume that $x^0 = ct$, so $dx^0 = c \, dt$, and the remaining three coordinates $x^i$, $i = 1, 2, 3$ will then represent spatial coordinates, in one sense of the work "spatial."

With Møller's terminology [**2**], a local frame field $\mathbf{e}_\mu$ will be called *time-orthogonal* iff:

$$g_{0i} = g(\mathbf{e}_0, \mathbf{e}_i) = 0 \qquad\qquad (i = 1, 2, 3). \tag{5.3}$$

In such a case, if $\theta^\mu$ is the reciprocal coframe field to $\mathbf{e}_\mu$ [so $\theta^\mu(\mathbf{e}_\nu) = \delta^\mu_\nu$], and we assume that $\theta^0$ is time-like, then we can express $g$ in the form:

$$g = (\theta^0)^2 - \gamma, \tag{5.4}$$

in which we have defined:

$$\gamma_s = -g_{ij}\,\theta^i\,\theta^j. \tag{5.5}$$

Hence, in that form, we can regard $T(U)$ as a direct sum $[\mathbf{e}_0](U) \oplus \Sigma(U)$ of the "time plus space" type, so that $g$ will take the form of the direct sum of a metric $(\theta^0)^2$ on $[\mathbf{e}_0](U)$ and another one $\gamma_s = \gamma_{ij}\,\theta^i\,\theta^j$ on $\Sigma(U)$.

Now, as is shown in Møller, if a metric is given in the general local form (5.2) then one can always find a suitable 1-form $\theta^0$ such that the local coframe field $\{\theta^0,\,dx^1,\,dx^2,\,dx^3\}$ will become time-orthogonal. Basically, one essentially "completes the square" that is defined by the first two terms in the expansion of $g$ as:

$$g = g_{00}\,(dx^0)^2 + 2g_{0i}\,dx^0\,dx^i + g_{ij}\,dx^i\,dx^j. \tag{5.6}$$

The solution to the problem is then:

$$\theta^0 = \frac{g_{0\mu}}{\sqrt{g_{00}}}\,dx^\mu = c^*\,dt + \gamma_i\,dx^i, \qquad \gamma_{ij} = \gamma_i\,\gamma_j - g_{ij}, \tag{5.7}$$

in which we have defined:

$$c^* = c\,\sqrt{g_{00}}, \qquad \gamma_i = \frac{g_{0i}}{\sqrt{g_{00}}}. \tag{5.8}$$

We shall refer to $\theta^0$ as the *temporal 1-form* that is associated with $g$ (and the choice of coordinate chart).

The association of the components of $g$ with the set of components $\{c^*,\,\gamma_i,\,\gamma_{ij}\}$ is a one-to-one correspondence whose inverse is:

$$g_{00} = \left(\frac{c^*}{c}\right)^2, \quad g_{0i} = g_{i0} = \frac{c^*}{c}\gamma_i, \qquad g_{ij} = \gamma_i\,\gamma_j - \gamma_{ij}. \tag{5.9}$$

Note that $c/c^* = 1/\sqrt{g_{00}}$ has the character of an "index of refraction" for the space at each point. Indeed, $g$ also defines a metric on $T^*M$, along with one on $T(M)$, namely:

$$g = g^{\mu\nu}\,\partial_\mu\,\partial_\nu = \frac{1}{c^2}\,g^{00}(\partial_t)^2 + \frac{2}{c}\,g^{0i}\partial_t\partial_i + g^{ij}\partial_i\partial_j. \tag{5.10}$$

The components of the metric on $T^*M$ will be $g^{\mu\nu}$, which is the inverse matrix to $g_{\mu\nu}$, so one will find that when the same basic time-orthonormalization process is applied to $g^{\mu\nu}$, one will get:

$$\overline{\mathbf{e}}_0 = \frac{g^{0\mu}}{\sqrt{g^{00}}}\,\partial_\mu = \frac{1}{c^{**}}\partial_t + \gamma^i\,\partial_i, \qquad \gamma^{ij} = \gamma^i\,\gamma^j - g^{ij}, \tag{5.11}$$

with:

$$c^{**} = \frac{c}{\sqrt{g^{00}}}, \qquad \gamma^i = \frac{g^{0i}}{\sqrt{g^{00}}}. \qquad (5.12)$$

Hence, in this case:

$$\sqrt{g^{00}} = \frac{c}{c^{**}}, \qquad (5.13)$$

which is more akin to an index of refraction.

It is clear from the second equation in (5.7) that the natural coframe field $dx^\mu$ will be time-orthogonal iff $\gamma_i = 0$, which would also make $\theta^0 = c^* \, dt$ and $\gamma_{ij} = -g_{ij}$. It is also clear that the projection of $g$ onto the spatial hyperplane $\Sigma$ that is normal to $\theta^0$ is not merely achieved by singling out the spatial components $g_{ij}$ of $g$.

We illustrate the relationship between $\Sigma$, $dt$, and $\theta^0$ in Fig. 1.

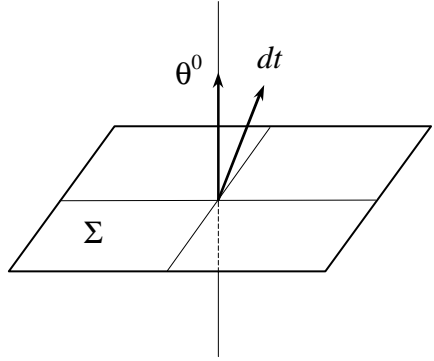

Figure 1. Time-orthogonalization of a general metric.

If one makes the substitutions:

$$g = c^2 \, (d\tau)^2, \quad dx^0 = c \, dt, \qquad dx^i = \frac{dx^i}{dt} \, dt = v^i \, dt, \qquad (5.14)$$

which correspond to evaluating $g$ along a world-line that is parameterized by proper time $\tau$ in four dimensions and by the time coordinate $t$ in the three-dimensional space then one will get:

$$c^2 \, (d\tau)^2 = [g_{00} \, c^2 + 2g_{0i} \, c \, v^i + g_{ij} \, v^i \, v^i] \, (dt)^2,$$

and with the substitution $dt = (dt \, / \, d\tau) \, d\tau$, this will give:

$$\frac{dt}{d\tau} = \left[ g_{00} + 2g_{0i} \, \frac{v^i}{c} + \frac{g_{ij} v^i v^j}{c^2} \right]^{-1/2}, \qquad (5.15)$$

which will become the Fitzgerald-Lorentz contraction factor when $g = \eta$.

We can also express this in terms of the time-orthogonal component form of $g$ by means of the substitution (5.9):

$$\frac{dt}{d\tau} = \frac{c}{c^*}\left[\left(1 + \frac{\gamma_i v^i}{c^*}\right)^2 - \left(\frac{v}{c^*}\right)^2\right]^{-1/2} \qquad (v^2 = \gamma_{ij}\, v^i\, v^j). \qquad (5.16)$$

If $g$ were time-orthogonal to begin with ($\gamma_i = 0$, $\gamma_{ij} = -g_{ij}$) then this would reduce to something that would be proportional to the Fitzgerald-Lorentz factor, but with a rescaling of $c$ to $c^*$ by means of $\sqrt{g_{00}}$. Note that for a general $g$ (i.e., $g_{00} \neq 1$), even in the rest space ($v^i = 0$), $dt$ will not be equal to $d\tau$, but only proportional, by way of $1/\sqrt{g_{00}}$.

   *b. Physical interpretation of* $\theta^0$. – In order to interpret $\theta^0$ physically, it helps to first examine the way that space-time geodesics of the metric $g$ turn into spatial geodesics, as viewed in the "instantaneous rest space." The form that the four-dimensional geodesic equation takes when one starts with the variational problem (viz., the Euler-Lagrange equations) is:

$$\frac{d}{d\tau}\left(g_{\mu\nu} u^\nu\right) = \tfrac{1}{2}\partial_\mu g_{\lambda\nu}\, u^\lambda\, u^\nu \qquad \left(u^\mu = \frac{dx^\mu}{d\tau}\right), \qquad (5.17)$$

when the curve parameter is proper time $\tau$. Although this will take on the Levi-Civita form when one performs the differentiation on the left-hand side, symmetrizes it with respect to $\lambda$, $\nu$, and rearranges terms, this form is more useful at the moment.

   Since only the spatial equations are of immediate interest to us, we set $\mu = i$ and get:

$$\frac{d}{d\tau}\left(g_{i0}\, u^0\right) + \frac{d}{d\tau}\left(g_{ij} u^j\right) = \tfrac{1}{2}[\partial_i g_{00}\,(u^0)^2 + 2\partial_i g_{0j}\, u^0\, u^j + \partial_i g_{jk}\, u^j\, u^k].$$

If one expands the first term on the left then one will get:

$$\partial_0 g_{i0}\,(u^0)^2 + \partial_j g_{i0}\, u^0 u^j + g_{i0}\,\frac{du^0}{d\tau} + \partial_k g_{ij} u^j u^k + g_{ij}\,\frac{du^j}{d\tau}$$
$$= \tfrac{1}{2}[\partial_i g_{00}\,(u^0)^2 + 2\partial_i g_{0j}\, u^0\, u^j + \partial_i g_{jk}\, u^j\, u^k].$$

In the instantaneous rest space, one will have $u^0 = c\, dt/d\tau$, $u^i = 0$, and the last equation will reduce to:

$$c^2 \partial_0 g_{i0}\left(\frac{dt}{d\tau}\right)^2 + g_{i0}\, c\,\frac{dt}{d\tau} + g_{ij}\,\frac{du^j}{d\tau} = \tfrac{1}{2}c^2\left(\frac{dt}{d\tau}\right)^2 \partial_i g_{00}. \qquad (5.18)$$

   Now:

$$\frac{du^j}{d\tau} = \frac{d}{d\tau}\left(\frac{dt}{d\tau} v^j\right) = \frac{d^2 t}{d\tau^2} v^j + \left(\frac{dt}{d\tau}\right)^2 a^j \qquad \left(a^j \equiv \frac{dv^j}{dt}\right),$$

which will become:

$$\frac{du^j}{d\tau} = \left(\frac{dt}{d\tau}\right)^2 a^j = \left(\frac{c}{c^*}\right)^2 a^j$$

in the instantaneous rest space, and that will put (5.18) into the form:

$$\left(\frac{c}{c^*}\right)^2 c^2 \partial_0 g_{i0} + g_{i0}\, c\, \frac{d^2 t}{d\tau^2} + \left(\frac{c}{c^*}\right)^2 g_{ij} a^j = \tfrac{1}{2} c^2 \left(\frac{c}{c^*}\right)^2 \partial_i g_{00}\,.$$

If we recall (5.16) then we will get:

$$\frac{d^2 t}{d\tau^2} = \frac{d}{d\tau}\left(\frac{c}{c^*}\right)\left[\left(1+\frac{\gamma_i v^i}{c^*}\right)^2 - \left(\frac{v}{c^*}\right)^2\right]^{-1/2} + \frac{c}{c^*}\frac{d}{d\tau}\left[\left(1+\frac{\gamma_i v^i}{c^*}\right)^2 - \left(\frac{v}{c^*}\right)^2\right]^{-1/2}$$

$$= -\left(\frac{c}{c^*}\right)^2 \frac{1}{c^*}\left(\partial_t c^* + \gamma_j a^j\right)$$

in the instantaneous rest space.  Substituting that into the previous equation, along with the expressions for $g_{00}$, $g_{0i}$, $g_{ij}$ in terms of $c^*$, $\gamma_i$, $\gamma_{ij}$ will give:

$$\gamma_{ij}\, a^j = c^*(\partial_t \gamma_i - \partial_i\, c^*)\,, \tag{5.19}$$

which agrees with (8.107) in Møller, up to our difference in sign convention for the components of $g$.

Hence, we have obtained an expression for the spatial acceleration, as measured in the instantaneous rest frame.  Since it is a function of only the components of the space-time metric, one sees that the acceleration in question can only be gravitational in origin.  Therefore, the components $a^i$ could just as well serve as the components of the spatial vector field $\mathbf{g}$ that one usually employs in Newtonian gravitation.

By comparison, if we compute the exterior derivative of $\theta^0$ then we will get:

$$d_\wedge \theta^0 = d_\wedge c^* \wedge dt + d_\wedge \gamma_i \wedge dx^i = \partial_i c^* dx^i \wedge dt + \partial_t \gamma_i\, dt \wedge dx^i + \partial_j \gamma_i\, dx^j \wedge dx^i\,,$$

which will take on the ultimate form:

$$d_\wedge \theta^0 = (\partial_t \gamma_i - \partial_i c^*)\, dt \wedge dx^i + \tfrac{1}{2}(\partial_i \gamma_j - \partial_j \gamma_i)\, dx^i \wedge dx^j\,. \tag{5.20}$$

One then sees from (5.19) that the covariant components (in the instantaneous rest frame) of the spatial acceleration due to gravity $a^i$ are simply $c^*$ times the time-space components of the 2-form $d_\wedge\theta^0$.  Since one also has that the exterior derivative $d_\wedge u$ of a covelocity 1-form $u$ will have the spatial acceleration for its time-space components, one can say that $c^* d_\wedge \theta^0$ is the kinematical vorticity of some covelocity 1-form that relates to $\theta^0$.

For instance, $c^* \theta^0$ would have $a^i$ for the time-space components of its kinematical vorticity if $c^*$ (and therefore $g_{00}$) were constant in time and space. More generally, if $c^*$ were constant in time then the time-space components of $c^* d \wedge \theta^0$ would take the form $\partial_t(c^* \gamma_i) - \partial_i(\frac{1}{2} c^{*2})$, which would be consistent $d \wedge \phi$, with:

$$\phi = \tfrac{1}{2} c^{*2} dt + c^* \gamma_i \, dx^i = c^* \theta^0 - \tfrac{1}{2} c^{*2} \, dt \,, \qquad (5.21)$$

which is not equal to $c^* \theta^0$ precisely.

*c. Integrability of the splitting.* – Since $dx^1$, $dx^2$, $dx^3$ are already completely-integrable (viz., exact), the only real issue that bears upon the degree of integrability of the coframe $\{\theta^0, dx^1, dx^2, dx^3\}$ is the degree of integrability of the exterior differential system $\theta^0 = 0$. That, in turn, depends upon the vanishing (or not) of the successive integrability $k$-forms that $\theta^0$ defines. In order to calculate them explicitly, we begin by recalling the definition of $\theta^0$ in (5.7) and (5.8):

$$\theta^0 = \frac{g_{0\mu}}{\sqrt{g_{00}}} dx^\mu = c^* dt + \gamma \qquad (c^* = c\sqrt{g_{00}} \,, \ \gamma = \gamma_i \, dx^i, \ \gamma_i = g_{0i} / \sqrt{g_{00}} \,). \quad (5.22)$$

We then get:

$$\Omega \equiv d \wedge \theta^0 = dt \wedge a_r + \omega_r \equiv dt \wedge (\partial_t \gamma - d_s c^*) + d_s \wedge \gamma, \qquad (5.23)$$

$$\theta^0 \wedge \Omega = dt \wedge (c^* \omega_r - \gamma \wedge a_r) + \gamma \wedge \omega_r, \qquad (5.24)$$

$$\Omega \wedge \Omega = -2 \, dt \wedge \omega_r \wedge a_r. \qquad (5.25)$$

In this, we have recalled the definitions of $a_r$ and $\omega_r$ in (4.6) and (4.7) that grew out (4.4) in order to emphasize the formal analogy that comes about by treating the timelike 1-form $\theta^0$ as if it were a relativistic covelocity 1-form.

The vanishing of the three differential forms $\Omega$, $\theta^0 \wedge \Omega$, $\Omega \wedge \Omega$ will give:

$$0 = a_r, \qquad\qquad 0 = \omega_r, \qquad (5.26)$$

$$0 = c^* \omega_r - \gamma \wedge a_r, \qquad 0 = \gamma \wedge \omega_r, \qquad (5.27)$$

$$0 = \omega_r \wedge a_r, \qquad (5.28)$$

respectively.

One should first note that the problem of establishing the degree of integrability of $\theta^0 = 0$ subsumes the problem of establishing the degree of integrability of the purely spatial Pfaff equation $\gamma_i = 0$; i.e., the vanishing of $\Omega_s$ and $\gamma \wedge \Omega_s$. (The vanishing of the 4-form $\Omega_s \wedge \Omega_s$ is an identity.) Those last conditions are necessary for $\gamma$ to take the forms $dv$ and $\mu \, dv$, respectively; otherwise, $\gamma$ will take the form $d\lambda + \mu \, dv$, which is analogous to the introduction of Clebsch variables for vortical flows in hydrodynamics [3, 10].

The first condition (5.26) is necessary for $\theta^0$ to take the form:

$$\theta^0 = c \, dt'. \tag{5.29}$$

In effect, that would make $\theta^0$ a coordinate differential for the coordinate $ct'$, so the spatial hypersurfaces would be level hypersurfaces of $t'$. When one expands the differential in (5.29) into temporal and spatial parts, one will get:

$$\theta^0 = c \, \frac{\partial t'}{\partial t} dt + c \, d_s t'. \tag{5.30}$$

Of course, if the space-time manifold is simply-connected then the condition (5.26) will also be sufficient for $\theta^0$ to take the form (5.29).

If one interprets $\theta^0$ as a timelike covelocity 1-form then:

**Theorem:**

1. *If the flow of the vector field that corresponds to $\theta^0$ is unaccelerated and spatially irrotational then $\theta^0$ can be used as the differential of a time coordinate $ct'$.*

2. *The spatial manifolds will then be the level hypersurfaces of $t'$.*

3. *One will have:*

$$c^* = c \, \partial_t t', \qquad \gamma = c \, d_s t'. \tag{5.31}$$

The second condition (5.27) is necessary for $\theta^0$ to take the form:

$$\theta^0 = c' \, dt', \tag{5.32}$$

in which $c'$ is a differentiable function now. In this case, $\theta^0$ cannot be a coordinate differential, but it is proportional to one, so the spatial hypersurfaces will again be level hypersurfaces of the function $t'$. Splitting (5.32) into temporal and spatial parts will give:

$$\theta^0 = c' \frac{\partial t'}{\partial t} dt + c' \, d_s t'. \tag{5.33}$$

Hence:

**Theorem:**

1. *If the flow of the vector field that corresponds to $\theta^0$ is hypersurface-orthogonal then $\theta^0$ will take the form of an integrating factor $c'$ times the differential of a time coordinate $t'$.*

2. *The spatial submanifolds will then be the level hypersurfaces of $t'$.*

3. *One will have:*

$$c^* = c' \, \partial_t t', \qquad \gamma = c' \, d_s t'. \tag{5.34}$$

The third condition (5.28) is necessary for $\theta^0$ to take the form:

$$\theta^0 = c \, dt' + c'' \, dt'', \tag{5.35}$$

in which $c$ is, of course, constant. When $\theta^0$ takes this form, the spatial submanifolds will no longer be three-dimensional, but two-dimensional, and will take the form of the intersections of the level-hypersurfaces of the functions $t'$ and $t''$. Splitting (5.35) into temporal and spatial parts gives:

$$\theta^0 = (c \, \partial_t t' + c'' \, \partial_t t'') \, dt + c \, d_s t' + c'' \, d_s t''. \tag{5.36}$$

Hence:

**Theorem:**

1. *If the flow of the vector field that corresponds to $\theta^0$ has its linear acceleration tangential to its vortex lines then $\theta^0$ will take the form* (5.35).

2. *The spatial submanifolds will then be the intersections of the level hypersurfaces of $t'$ and those of $t''$, and will therefore be two-dimensional.*

3. *One will have:*

$$c^* = c \, \partial_t t' + c'' \, \partial_t t'', \qquad \gamma = c \, d_s t' + c'' \, d_s t''. \tag{5.37}$$

Finally, if $\theta^0$ takes the form:

$$\theta^0 = c' \, dt' + c'' \, dt'' \tag{5.38}$$

then $\Omega \wedge \Omega$ will not vanish, and neither will $\omega_r \wedge a_r$. The splitting of (5.38) into time + space will be:

$$\theta^0 = (c' \, \partial_t t' + c'' \, \partial_t t'') \, dt + c' \, d_s t' + c'' \, d_s t''. \tag{5.39}$$

Hence:

**Theorem:**

1. *If the flow of the vector field that corresponds to $\theta^0$ has its linear acceleration transverse to its vortex lines then $\theta^0$ will take the form* (5.38).

2. *The spatial submanifolds will then be the intersections of the level hypersurfaces of $t'$ and those of $t''$, and will therefore be two-dimensional.*

3. *One will have:*

$$c^* = c' \partial_t t' + c'' \, \partial_t t'', \qquad \gamma = c' d_s t' + c'' \, d_s t''. \tag{5.40}$$

**6. Physical examples.** – We shall now examine some of the physical examples of space-time geometries in regard to the degree of integrability of their temporal 1-form $\theta^0$.

*a. Stationary space-times.* – A space-time is called *stationary* when there exists some coordinate system in which its metric is time-orthogonal and thus takes the form:

$$g = g_{00} \, c^2 \, (dt)^2 - g_{ij} \, dx^i \, dx^j. \tag{6.1}$$

Hence, in that coordinate system, one must have:

$$\theta^0 = c^* \, dt, \qquad \gamma_i = 0. \tag{6.2}$$

The form of $\theta^0$ for a stationary space-time is that of an exact form with an integrating factor. Hence:

**Theorem:**

*Any stationary space-time has a completely-integrable temporal 1-form $\theta^0$.*

Therefore, the space-time manifold can be foliated by three-dimensional leaves that amount to the level hypersurfaces of $t$; i.e., simultaneity hypersurfaces of the time coordinate.

Notice that the vorticity of $\theta^0$ can still be non-zero, but since:

$$\Omega = d_s c^* \wedge dt, \tag{6.3}$$

it will be indifferent to the time-dependency of $c^*$.

Although this particular example seems somewhat restricting on first glance, nonetheless, it includes a large number of the traditional solutions of the Einstein field equations, such as those of Schwarzschild, Robertson-Walker, de Sitter and anti-de Sitter, and the Reissner-Nordstrøm solution.

A stationary space-time should not be confused with a *static* space-time, which is one for which there exists a coordinate system in which the components $g_{\mu\nu}$ of $g$ are time-independent; $\partial_t \, g_{\mu\nu} = 0$, which will imply that $\partial_t \, c^*$ and $\partial_t \, \gamma$ will also have to vanish. For such a space-time, one can simplify the integrability forms somewhat to:

$$\Omega \qquad = - dt \wedge d_s c^* + d_s \wedge \gamma, \tag{6.4}$$
$$\theta^0 \wedge \Omega = - dt \wedge \gamma \wedge d_s c^* + \gamma \wedge d_s \wedge \gamma, \tag{6.5}$$
$$\Omega \wedge \Omega = - 2 \, dt \wedge d_s c^* \wedge d_s \wedge \gamma. \tag{6.6}$$

*b. Rotating disc.* – The rotating disc gives an elementary, but fascinating, example of a geometry for a surface that is static, but not stationary. Moreover, it can be shown to be a surface with negative Gaussian curvature.

One starts by defining the local diffeomorphism from $\mathbb{R}^3$ with coordinates $(t, r, \theta)$, to $\mathbb{R}^3$ with coordinates $(\bar{t}, x, y)$:

$$\left.\begin{aligned} \bar{t} &= t, \\ x &= r\cos(\theta + \omega t), \\ y &= r\sin(\theta + \omega t), \end{aligned}\right\} \tag{6.7}$$

which differentiates to:

$$\left.\begin{aligned} d\bar{t} &= \ dt, \\ dx &= -\omega r\cos dt + \cos dr - r\sin d\theta, \\ dy &= \ \omega r\sin dt + \sin dr + r\cos d\theta. \end{aligned}\right\} \tag{6.8}$$

With those substitutions, the three-dimensional Minkowski space metric:

$$\eta = c^2\,(d\bar{t})^2 - (dx)^2 - (dy)^2 \tag{6.9}$$

will pull back to:

$$\eta = c^2\left[1 - \left(\frac{\omega r}{c}\right)^2\right](dt)^2 - 2\omega r^2\,dt\,d\theta - (dr)^2 - r^2\,(d\theta)^2. \tag{6.10}$$

That will make:

$$c^* = c\left[1 - \left(\frac{\omega r}{c}\right)^2\right]^{1/2}, \qquad \gamma = -\frac{\omega r^2}{\sqrt{1 - \left(\frac{\omega r}{c}\right)^2}}\,d\theta. \tag{6.11}$$

Hence:

$$\theta^0 = c\,\sqrt{1 - \left(\frac{\omega r}{c}\right)^2}\,dt + \frac{\omega r^2}{\sqrt{1 - \left(\frac{\omega r}{c}\right)^2}}\,d\theta = c^*(r)\,dt + \frac{c}{c^*}\,\omega r^2\,d\theta. \tag{6.12}$$

This has the normal form $\mu_1\,(r)\,dv_1 + \mu_2\,(r)\,dv_2$, so we expect that the integral submanifolds of $\theta^0$ in our three-dimensional space will be integral curves that represent the intersections of the level surfaces of $t$ and $\theta$; i.e., radial lines in the discs.

Just to check, we compute:

$$\Omega = -\frac{\omega r}{c^*}\,dr \wedge \left\{\omega\,dt + c\left[2 + \left(\frac{\omega r}{c}\right)^2\right]d\theta\right\}, \tag{6.13}$$

$$\theta^0 \wedge \Omega = -2c\omega r\,dt \wedge dr \wedge d\theta. \tag{6.14}$$

The latter expression will vanish only when $\omega$ or $r$ vanishes, although strictly speaking, $r = 0$ is not an allowable point.

One should note that for a given $\omega$, there will be a maximum radius $r_{\max}$ for the disc that is defined by the radius at which the tangential speed equals $c$; i.e., $r_{\max} = c / \omega$. Similarly, for a given radius $r$ there will be a maximum $\omega$, namely, $c / r$.

The spatial metric that is associated with the space-time metric (6.10) is:

$$\gamma_s = (dr)^2 + \frac{1}{1-(\omega r / c)^2} (r\, d\theta)^2, \tag{6.15}$$

which differs from the Euclidian metric by the factor that multiplies $(r\, d\theta)^2$. As $\omega$ goes to 0, that factor will converge to 1, which is the Euclidian case, but as $\omega r$ goes to $c$, it will become infinite.

*c. Uniformly-accelerated observer.* – Now, let a local diffeomorphism of the plane $\mathbb{R}^2 = (t, x)$ be defined by:

$$\overline{t} = t, \quad \overline{x} = x + \tfrac{1}{2}at^2, \tag{6.16}$$

in which $a$ is a constant.

This differentiates to:

$$d\overline{t} = dt, \qquad d\overline{x} = at\, dt + dx, \tag{6.17}$$

and the pull-back of the two-dimensional Minkowski space metric $c^2(d\overline{t})^2 - (d\overline{x})^2$ will take the form:

$$g = c^2 \left[ 1 - \left(\frac{at}{c}\right)^2 \right] (dt)^2 - 2at\, dt\, dx - (dx)^2. \tag{6.18}$$

Hence:

$$g_{00} = \left[ 1 - \left(\frac{at}{c}\right)^2 \right], \qquad g_{01} = g_{10} = \frac{at}{c}, \qquad g_{11} = 1, \tag{6.19}$$

which will make:

$$c^* = c\sqrt{1 - \left(\frac{at}{c}\right)^2}, \qquad \gamma = \frac{at}{c^*}\, dx, \tag{6.20}$$

and

$$\theta^0 = c\sqrt{1 - \left(\frac{at}{c}\right)^2}\, dt + \frac{at/c}{\sqrt{1-(at/c)^2}}\, dx, \tag{6.21}$$

Since space-time is two-dimensional, $\theta^0$ will have either $dv$ or $\mu\, dv$ for its normal form, so it must be completely-integrable, in either case.

However, if one computes $\Omega$:

$$\Omega = \frac{a/c}{[1-(at/c)^2]^{3/2}}\, dt \wedge dx \tag{6.22}$$

then one will see that this vanishes only as $t$ becomes infinite; hence, $\theta^0$ cannot be exact.

"Space" in this case is one-dimensional (i.e., the $x$-axis), and the spatial metric that is associated with $g$ will take the form:

$$\gamma = \frac{1}{1 - (at/c)^2} (dx)^2. \tag{6.23}$$

Hence, one will find a gradual contraction of lengths up to the limiting value of time at which $at = c$.

Unlike the uniform rotation, which changes the curvature of "space," the uniform linear acceleration will not. In essence, that is because the space of a rotating disc is at least two-dimensional, while the space of a uniform acceleration is one-dimensional, which must necessarily have vanishing curvature.

*d. Gödel space-time.* – In 1949, Kurt Gödel published an exact solution to the Einstein field equations [**16**] that represented a somewhat-unphysical, but still instructive, cosmology in which the cosmic dust distribution was rotating (rigidly and collectively) about the $z$-axis with a constant angular velocity of $\omega$. The metric that he obtained was:

$$g = \frac{1}{2\omega^2} [(c\, dt + e^{x/x_0}\, dy)^2 - (dx)^2 - \tfrac{1}{2} e^{2x/x_0} (dy)^2 - (dz)^2]. \tag{6.24}$$

Since we are only concerned with the integrability of its time-orthogonal temporal 1-form $\theta^0$, we shall drop the leading constant factor of $1/2\omega^2$. Furthermore, since the $z$-axis is clearly passive to this metric, we drop the $(dz)^2$ from the expression for $g$ and consider a variation on the rotating-disc geometry.

This metric is already in time-orthogonal form, so we can read off:

$$\theta^0 = c\, dt + e^{x/x_0}\, dz . \tag{6.25}$$

This is already in the normal form for a 1-form on a three-dimensional manifold that is not completely-integrable. Hence, the integral submanifolds of $\theta^0 = 0$ will be the intersections of the level surfaces for $t$ and $z$; i.e., the $y$-axis at each value of $t$.

Just to check, we compute:

$$\Omega = \frac{1}{x_0} e^{x/x_0}\, dx \wedge dz, \tag{6.26}$$

$$\theta^0 \wedge \Omega = \frac{c}{x_0} e^{x/x_0}\, dt \wedge dx \wedge dz. \tag{6.27}$$

Naturally, $\Omega \wedge \Omega = 0$, since our space-time is only three-dimensional.

The first of these equations shows that $\theta^0$ has non-vanishing vorticity, while the second one shows that the integral curves of the vector field that corresponds to $\theta^0$ are not hypersurface-orthogonal.

*e. Kerr space-time.* – In 1961, Roy Kerr found an exact solution of the Einstein equations for the case in which the source mass distribution is static, spherically-symmetric and rotating [**17**]. Hence, it subsumed the Schwarzschild solution as the case in which the angular velocity is zero, and is commonly used to represent rotating black holes nowadays. In 1965, Ezra Newman extended that solution to the case of a charged, rotating, spherically-symmetric mass [**18**]. We shall examine only the former case, which is already quite involved.

The Kerr metric takes the form:

$$g = \left(1 - \frac{r_S\, r}{\rho^2}\right) c^2 dt^2 + 2\frac{r_S\, r \alpha \sin^2 \theta}{\rho^2} c\, dt\, d\phi - \frac{\rho^2}{\Delta}(dr)^2 - \rho^2 (d\theta)^2$$
$$- \left(r^2 + \alpha^2 + \frac{r_S\, r \alpha}{\rho^2} \sin^2 \theta\right) \sin^2 \theta (d\phi)^2. \qquad (6.28)$$

In this expression, we have introduced the Schwarzschild radius $r_S = 2GM / c^2$ ($M$ = mass of stellar object) and the length scales:

$$\alpha = \frac{J}{Mc}, \qquad \rho^2 = r^2 + \alpha^2 \cos^2 \theta, \quad \Delta = r^2 - r_S\, r + \alpha^2,$$

in which $J$ is the angular momentum of the rotating mass. When $J = 0$, we will have $\alpha = 0$, $\rho = r$, and $\Delta = r^2 - r_S\, r$, which will make:

$$g(J = 0) = \left(1 - \frac{r_S}{r}\right) c^2 dt^2 - \left(1 - \frac{r_S}{r}\right)^{-1} (dr)^2 - r^2[(d\theta)^2 + \sin^2 \theta (d\phi)^2], \qquad (6.29)$$

which is indeed the Schwarzschild metric.

We can read off directly from (6.28):

$$c^* = c\sqrt{1 - \frac{r_S\, r}{\rho^2}}, \qquad \gamma = \frac{r_S\, r \alpha \sin^2 \theta}{\rho^2 \sqrt{1 - r_S\, r / \rho^2}} d\phi. \qquad (6.30)$$

Hence:

$$\theta^0 = c\sqrt{1 - \frac{r_S\, r}{\rho^2}}\, dt + \frac{r_S\, r \alpha \sin^2 \theta}{\rho^2 \sqrt{1 - r_S\, r / \rho^2}} d\phi. \qquad (6.31)$$

This corresponds to the normal form $\mu_1(r, \theta)\, dv_1 + \mu_1(r, \theta)\, dv_1$, which suggests that the integral submanifolds of $\theta^0 = 0$ are two-dimensional, and will correspond to the intersections of the level hypersurfaces of $t$ and $\phi$; hence, they will be the spatial cones that are described by $r$ and $\theta$ when $t$ and $\phi$ are held constant.

Since the exact expressions for $\Omega$, $\theta^0 \wedge \Omega$, and $\Omega \wedge \Omega$ are quite complicated and hardly illuminating, we shall omit those calculations.

**7. Summary.** – Let us summarize what we have accomplished in this article:

1. Any Lorentzian metric $g$ on a space-time manifold can be expressed locally in time-orthogonal form:

$$g = (\theta^0)^2 - \gamma_s$$

by a process of completing the square.

2. Whether or not the spatial hypersurfaces that are defined by $\theta^0 = 0$ are tangent to spatial hypersurfaces depends upon the degree of the integrability of that Pfaff equation.

3. The degree of integrability of the aforementioned Pfaff system can be ascertained by either putting $\theta^0$ into normal form or looking at the first vanishing $k$-form in the sequence:

$$d_\wedge \theta^0, \; \theta^0 \wedge d_\wedge \theta^0, \; d_\wedge \theta^0 \wedge d_\wedge \theta^0, \; \theta^0 \wedge d_\wedge \theta^0 \wedge d_\wedge \theta^0, \; \ldots,$$

which will truncate when $k$ exceeds the dimension of the manifold.

4. Any space-time metric that is already in time-orthogonal form has a temporal 1-form $\theta^0$ that is completely integrable. This category includes many of the most familiar exact solutions of the Einstein equations.

5. Some of the established exact solutions, such a those of Gödel, Kerr, and Newman, are not completely integrable and admit only two-dimensional spatial surfaces.

## References (*)

1.      D. H. Delphenich:
        – "Proper Time Foliations of Lorentz Manifolds," arXiv.org:gr-qc/0211066.
        – "Nonlinear connections and 1+3 splittings of spacetime," arXiv:gr-qc/0702115.
        – "Transverse geometry and physical observers," arXiv:0711.2033.
2.      C. Møller, *The Theory of Relativity*, 2$^{nd}$ ed., Oxford Univ. Press, 1972.
3.      D. H. Delphenich, "The role of integrability in a large class of physical systems," arXiv:1210.4976.
4.      W. Greub, *Multilinear Algebra*, Springer Verlag, Berlin, 1967.
5.*     E. Cartan, "Sur l'intégration des systèmes d'équations aux différentielles totales," Ann. Ec. Norm. Sup. **18** (1901), 241-311.
6.*     Kähler, E., *Einführung in die Theorie der Systeme von Differentialgleichungen,* Teubner, Leipzig, 1934.
7.      R. L. Bryant, S.-S. Chern, R. B. Gardner, H. L. Goldschmidt, P. A. Griffiths, *Exterior Differential Systems,* Springer, Berlin, 1991.

---

(*) References that are marked with an asterisk are available in English translation at the author's website neo-classical-physics.info.

8.*     E. Cartan, "Sur certaines expressions différentielles et le problème de Pfaff," Ann. Ec. Norm. Sup. **16** (1899), 239-332. English translation by D. H. Delphenich at neo-classical-physics.info.

9.     E. Goursat, *Leçons sur les problème de Pfaff*, Hermann, Paris, 1922.

10.     Sir H. Lamb, *Hydrodynamics*, $6^{th}$ *ed.*, Cambridge Univ. Press, Cambridge, 1932; (First ed. 1879), reprinted by Dover, Mineola, NY, 1945.

11.     D. Batchelor, *Introduction to Fluid Mechanics,* Cambridge University Press, Cambridge, 1967.

12.     J. L. Synge, "Relativistic Hydrodynamics," Proc. London Math. Soc. **43** (1937), 376-416.

13.*     A. Lichnerowicz:
    − *Théorie relativiste de la gravitational et de l'electromagnetisme,* Masson and Co., Paris, 1955.
    − *Relativistic hydrodynamics and magnetohydrodynamics*, W. A. Benjamin, Reading, MA, 1967.

14.     A. Anile, *Relativistic Fluids and Magnetofluids*, Cambridge Univ. Press, 2008.

15.     Fujimoto, A., *Theory of* G-*structures,* Publications of the Study Group of Geometry, 1972.

16.     K. Gödel, "An example of a new type of cosmological solution of Einstein's field equations of gravitation," Rev. Mod. Phys. **21** (3) (1949), 447–450.

17.     R. Kerr, "Gravitational Field of a Spinning Mass as an Example of Algebraically Special Metrics," Physical Review Letters **11** (5) (1963), 237–238.

18.     E. Newman, E. Couch, K. Chinnapared, A. Exton, A. Prakash, R. Torrence, "Metric of a Rotating, Charged Mass," Journal of Mathematical Physics **6** (6) (1965), 918–919.